\documentclass[preprint,superscriptaddress,longbibliography]{revtex4-2}

\usepackage{graphicx}
\usepackage{color}
\usepackage{amsfonts}

\usepackage{isomath}
\usepackage{amsmath}
\usepackage{amsthm}
\usepackage{amsfonts}
\usepackage{amssymb}
\usepackage{fdsymbol}
\usepackage{braket}
\usepackage{gensymb}

\let\oldAA\AA
\renewcommand{\AA}{\text{\normalfont\oldAA}}

\begin{document}
\title{Direct view of gate-tunable miniband dispersion in graphene superlattices near the magic twist angle}
\author{Zhihao Jiang}
\affiliation{Department of Physics and Astronomy, Interdisciplinary Nanoscience Center, Aarhus University, 8000 Aarhus C, Denmark}
\author{Dongkyu Lee}
\affiliation{Department of Physics, University of Seoul, Seoul 02504, Korea}
\affiliation{Department of Smart Cities, University of Seoul, Seoul 02504, Korea}
\author{Alfred J. H. Jones}
\affiliation{Department of Physics and Astronomy, Interdisciplinary Nanoscience Center, Aarhus University, 8000 Aarhus C, Denmark}
\author{Youngju Park}
\affiliation{Department of Physics, University of Seoul, Seoul 02504, Korea}
\author{Kimberly Hsieh}
\author{Paulina Majchrzak}
\author{Chakradhar Sahoo}
\author{Thomas S.  Nielsen}
\affiliation{Department of Physics and Astronomy, Interdisciplinary Nanoscience Center, Aarhus University, 8000 Aarhus C, Denmark}
\author{Kenji~Watanabe}
\affiliation{Research Center for Electronic and Optical Materials, National Institute for Materials Science, 1-1 Namiki, Tsukuba 305-0044, Japan}
\author{Takashi~Taniguchi}
\affiliation{Research Center for Materials Nanoarchitectonics, National Institute for Materials Science,  1-1 Namiki, Tsukuba 305-0044, Japan}
\author{Philip Hofmann}
\author{Jill A. Miwa}
\affiliation{Department of Physics and Astronomy, Interdisciplinary Nanoscience Center, Aarhus University, 8000 Aarhus C, Denmark}
\author{Yong P.  Chen}
\affiliation{Department of Physics and Astronomy, Interdisciplinary Nanoscience Center, Aarhus University, 8000 Aarhus C, Denmark}
\affiliation{Department of Physics and Astronomy and School of Electrical and Computer Engineering and Purdue Quantum Science and Engineering Institute, Purdue University, West Lafayette, IN 47907, USA\\
$^{\ast}$Email: ulstrup@phys.au.dk}
\author{Jeil Jung}
\affiliation{Department of Physics, University of Seoul, Seoul 02504, Korea}
\affiliation{Department of Smart Cities, University of Seoul, Seoul 02504, Korea}
\author{S{\o}ren Ulstrup$^{\ast}$}
\affiliation{Department of Physics and Astronomy, Interdisciplinary Nanoscience Center, Aarhus University, 8000 Aarhus C, Denmark}

\pagebreak

\begin{abstract}
Superlattices from twisted graphene mono- and bi-layer systems give rise to on-demand many-body states such as Mott insulators and unconventional superconductors. These phenomena are ascribed to a combination of flat bands and strong Coulomb interactions.  However, a comprehensive understanding is lacking because the low-energy band structure strongly changes when the electron filling is varied.  Here, we gain direct access to the filling-dependent low energy bands of twisted bilayer graphene (TBG) and twisted double bilayer graphene (TDBG) by applying micro-focused angle-resolved photoemission spectroscopy to \emph{in situ} gated devices. Our findings for the two systems are in stark contrast: The doping dependent dispersion for TBG can be described in a simple model, combining a filling-dependent rigid band shift with a many-body related bandwidth change. In TDBG, on the other hand, we find a complex behaviour of the low-energy bands, combining non-monotonous bandwidth changes and tuneable gap openings.  Our work establishes the extent of electric field tunability of the low energy electronic states in twisted graphene superlattices and can serve to underpin the theoretical understanding of the resulting phenomena.
\end{abstract}

\maketitle
\newpage

\section*{Introduction}
Stacking two graphene layers with a small interlayer twist angle $\theta$ gives rise to a long-range moir\'e structure and brings the linearly dispersing Dirac states at the Brillouin zone (BZ) corners within close vicinity of each other.  In twisted bilayer graphene (TBG), hybridization between the Dirac states leads to the formation of low energy bands in the resulting moir\'e mini Brillouin zone (mBZ). In contrast to the high-velocity Dirac states, these low-energy ``minibands'' consist of hole-like lower branches (LB) and electron-like upper branches (UB) that are essentially flat.  Additionally,  the opening of hybridization gaps separates the LB and UB from the continua of bands below the valence band maximum (VBM) and above the conduction band minimum (CBM) \cite{Bistritzer:2011,Santos:2012,Cao:2016},  as sketched in Fig.  \ref{fig:0}(a).  The electron filling in the flat minibands can be widely tuned by applying a vertical electric field.  The resulting changes in the carrier density, $n$, can shift the Fermi level in a hybridization band gap and induce an insulating state \cite{Cao:2016,Kyounghwan:2017}. As an alternative to graphene layers, a twisted heterostructure can also be formed by two Bernal-stacked bilayers of graphene.  In this material, twisted double bilayer graphene (TDBG), the minibands arise from  two hyperbolic bands with electric field-tunable band gaps (see  Fig.  \ref{fig:0}(b)) \cite{PhysRevB.99.235417,PhysRevB.99.235406,PhysRevB.100.201402,Haddadi2020,PhysRevLett.123.197702}, leveraging a wider field-tunability of the minibands \cite{Cao2020,Liu2020,Shen:2020,He2021}.  Near the ``magic'' twist angle,  $\theta \approx 1.1^{\circ}$,  the minibands become extremely flat such that the Coulomb interaction $U$ dominates over the kinetic energy,  as measured by the bandwidth $W$.  In this condition, half-integer filling levels of the minibands trigger correlated insulator states in both TBG and TDBG \cite{Cao:2018b,Cao2020}.

Due to the prominent role of many-body effects, changing the carrier density does not simply lead to a rigid shift of the minibands but rather to a complex change of the band dispersion. This is to be expected because the presence of van Hove singularities in the miniband dispersion can drastically affect the many-body effects when tuned to the Fermi level. The resulting spectral functions are extremely challenging to address theoretically  \cite{Guinea:2018,Rademaker:2019,Cea:2019,Goodwin:2020}. Experimentally, direct access to the doping-dependent miniband dispersion around the magic angle has not been achieved but signatures of miniband pinning due to van Hove singularities and symmetry-breaking cascade transitions have been observed in scanning tunneling spectroscopy experiments  \cite{Choi:2019,Jiang:2019,Yonglong:2019,Kerelsky:2019,Choi:2021}.  We gain direct access to the miniband $E(k)$-dispersion  by applying angle-resolved photoemission spectroscopy with micrometer spatial resolution (microARPES) to TBG and TDBG near the magic twist angle \cite{Utama:2021, Lisi:2021, Yiwei:2022,Nunn:2023, Jiang2023}. Crucially, the experiments of the present study are carried out in an optimum device geometry for photoemission spectroscopy,  such that we can resolve the detailed energy- and momentum-dependent spectral function at variable filling, and thereby directly access the many-body induced band deformations that result from long-range Coulomb interactions as the filling of minibands is tuned by a gate voltage \cite{Ulstrup:2016}. 

\section*{Results}

\begin{figure*}[t!] 
\begin{center}
\includegraphics[width=1.0\textwidth]{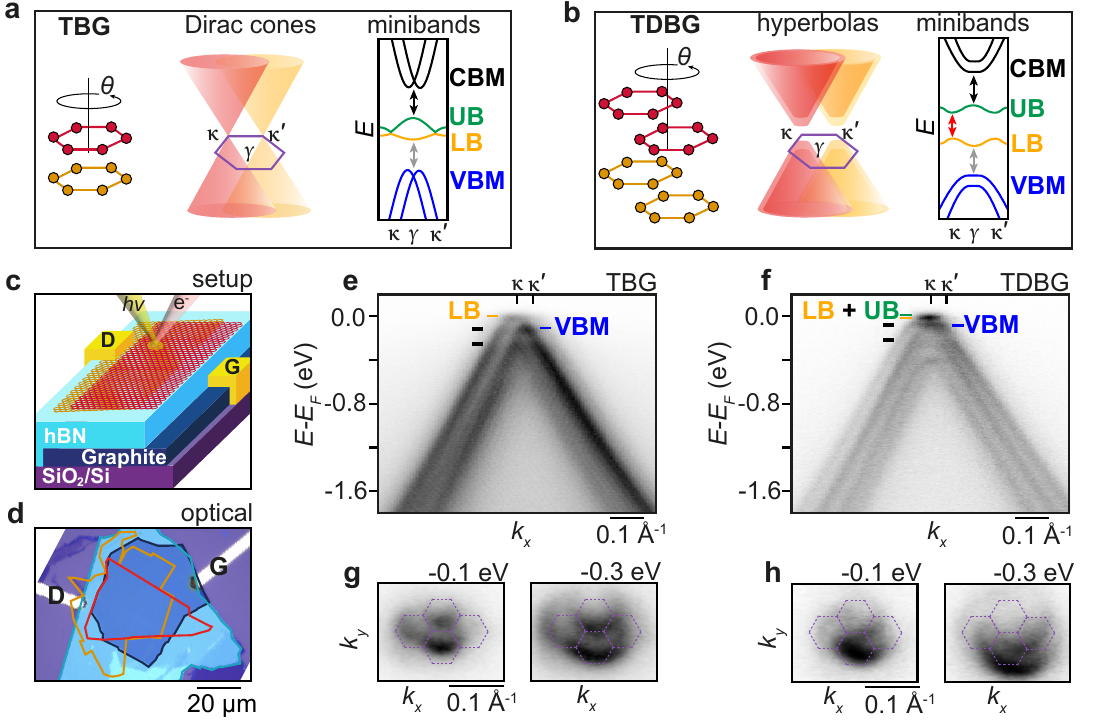}
\caption{Electronic structure of devices composed of twisted bilayer (TBG) and double bilayer graphene (TDBG) near the magic twist angle.  \textbf{a-b,} Outline of mBZs around overlapping (\textbf{a}) Dirac cones originating from top (red) and bottom (orange) graphene layers of TBG and (\textbf{b}) hyperbolic bands centred on top (red) and bottom (orange) Bernal-stacked graphene bilayers of TDBG.  The miniband dispersion is sketched in each case,  providing an overview of our definitions of CBM, UB, LB and VBM.  Band gaps are demarcated by double-headed arrows.  \textbf{c,} Schematic illustration of devices probed via micro-focused photoemission. \textbf{d,} Optical micrograph of TBG device. Top graphene, bottom graphene, hBN and graphite gate are demarcated by red, orange, light blue and dark blue outlines, respectively. Drain and gate electrodes are labelled D and G, respectively.  \textbf{e-f,} ARPES spectra of (\textbf{e}) TBG and (\textbf{f}) TDBG along the $\kappa$-$\kappa^{\prime}$-direction of the mBZ.  \textbf{g-h,} Constant energy surfaces of (\textbf{g}) TBG and (\textbf{h}) TDBG at -0.1 eV and -0.3 eV.  The extracted energies are marked by black ticks in (\textbf{e}) and (\textbf{f}).  The LB, UB and VBM are marked by orange, green and blue ticks, respectively.}
\label{fig:0}
\end{center}
\end{figure*}

\textbf{Electronic structure of TBG and TDBG devices.} A sketch and an optical micrograph of our TBG sample is presented in Figs.  \ref{fig:0}(c)-(d).  A graphite back-gate,  demarcated by a dark blue outline,  is placed on SiO$_2$/Si and encompasses the majority of the area with twisted graphene layers,  with each graphene monolayer indicated by orange and red outlines.  A hexagonal boron nitride (hBN) layer with thickness around 30 nm,  demarcated by a light blue outline, is sandwiched between graphite and graphene flakes.  The top layers are grounded via a drain (D) electrode while a voltage is applied via the gate (G) electrode to the graphite.  Our TDBG device has been constructed following the same architecture.  Detailed microARPES intensity maps and corresponding optical images for the TBG and TDBG devices are presented in Fig 5.  Both devices contain regions of bare monolayer and Bernal-stacked bilayer graphene directly above the gate, which we use to calibrate the carrier concentration $n$ induced by the gate voltage $V_G$ (see Figs.  6-7).  The twist angle in the devices is  $\theta = (1.2 \pm 0.2)^{\circ}$,  as verified by conductive tip atomic force microscopy (C-AFM) measurements (see Fig 8). 

Figures \ref{fig:0}(e)-(h) present  $E(k)$-dispersion cuts and constant energy surfaces for (e, g) TBG and (f,h) TDBG at zero gate voltage.  The dispersion cuts in panels (e) and (f) have been extracted along the high symmetry $\kappa$-$\kappa^{\prime}$-direction of the mBZ (see Figs.  \ref{fig:0}(a)-(b)).  In the case of TBG in panel (e),  we observe a faint flat band at the Fermi energy,  which we identify as the LB (see orange tick).  We can resolve the LB state straddling the Fermi energy despite the low initial carrier concentration, which is below our accuracy level of $1\cdot 10^{-11}$ cm$^{-2}$,  because our measurement is performed at room temperature,  leading to sufficient thermal excitation of the band. The VBM is identified below the LB at an energy of -0.1 eV (see blue tick).  The $E(k)$-dispersion of TDBG in panel (f) displays an intense flat state at the Fermi level, which we interpret as being composed of the adjacent LB and UB states (see orange and green ticks),  which initially appear gapless.  These bands are separated from segments of interacting hyperbolic bands that define the flat VBM (see blue tick).  The carrier density at zero gate voltage is determined to be $(1.4 \pm 0.3) \cdot 10^{12}$ cm$^{-2}$,  such that there is substantial occupation and thus photoemission intensity from both LB and UB straddling the Fermi energy.  The constant energy surfaces in panels (g) and (h) are extracted at -0.1 eV and -0.3 eV,  respectively,  cutting through segments of the VB states that are separated by a hybridization gap in both materials.  This leads to intensity being concentrated in the center of mBZs at -0.1 eV and spiralling outward towards the mBZ edges at -0.3 eV, providing a guide to construct the mBZs (see dotted hexagons) from the ARPES intensity.  In both cases we estimate a reciprocal moir\'e lattice vector length of $0.06$ \AA$^{-1}$,  which is consistent with the twist angle obtained from C-AFM. \\

\begin{figure*}[t!] 
\begin{center}
\includegraphics[width=1.0\textwidth]{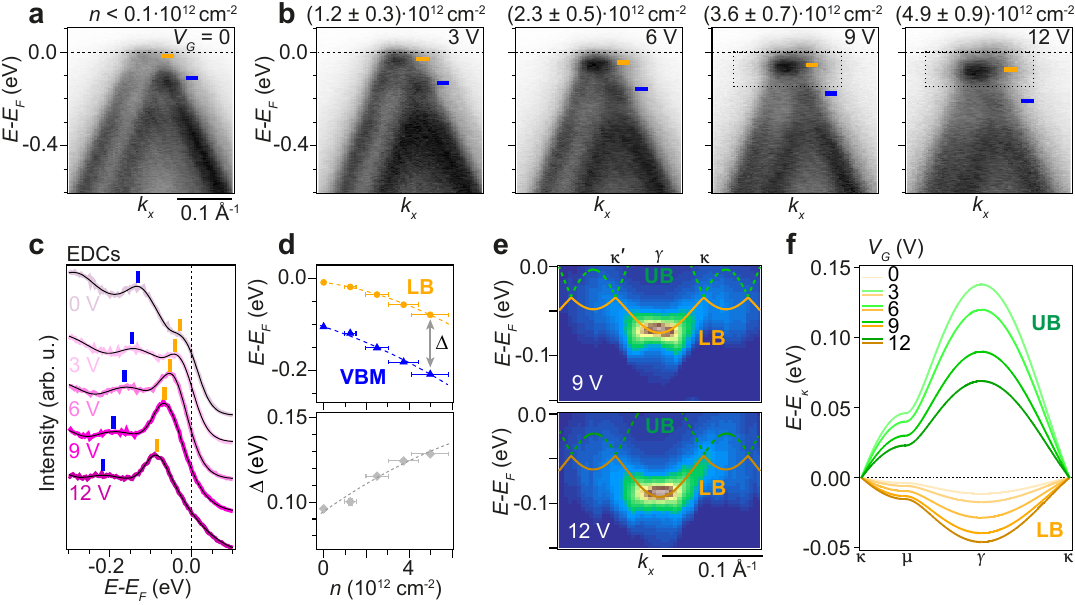}
\caption{Gate-tunable electronic structure of TBG.  \textbf{a-b,} ARPES spectra of TBG obtained at the stated values of gate voltage and carrier density.  \textbf{c,} Energy distribution curves (EDCs) integrated over a $k$-range of 0.04 \AA$^{-1}$ around the flat band region in (\textbf{a})-(\textbf{b}) and fits (smooth thin curves) to Lorentzian peaks multiplied by the Fermi-Dirac function at the stated gate voltages below the curves.  Fitted peak positions corresponding to LB and VBM are demarcated by orange and blue ticks,  respectively.  \textbf{d,} Top panel: Fitted peak positions from the EDC analysis in (\textbf{c}).  Bottom panel: Extracted values of the LB-VBM peak-to-peak energy difference $\Delta$.  The dashed lines represent polynomial fits that have been added as a guide to the eye.  \textbf{e,} Curvature plots of the ARPES intensity within the rectangular $(E,k)$-window outlined at 9 V and 12 V in (\textbf{b}). The overlaid LB and UB dispersion are results of a fit to the curvature intensity,  as described in Fig.  9 and Methods.  \textbf{f,} Fitted LB and UB dispersion relative to the $\kappa$-point energy at the measured filling factors.  Error bars on $n$ have been determined in connection with the doping calibration described in Fig 6.  The error bars on the EDC peak positions and $\Delta$ in (d),  that are derived based on the EDC fits, are smaller than the marker sizes. }
\label{fig:1}
\end{center}
\end{figure*}

\textbf{Gate tuning of bandwidth in TBG.} The gate-dependent dispersion of our TBG device is presented in Figs.  \ref{fig:1}(a)-(b).  The position of LB and the VBM are tracked in Fig. \ref{fig:1}(c) by fitting energy distribution curves (EDCs) integrated over a $k$-range of 0.04 \AA$^{-1}$,  encompassing LB and VB states. Note that this analysis gives the energy of highest photoemission intensity,  which coincides with the band position only for a truly flat band,  inducing some uncertainty in the case of dispersive states. As the gate voltage and carrier density are increased, LB  shifts to lower energy and becomes more intense (see orange tick mark that tracks the LB based on the EDC fits).  The VBM also shifts to lower energy and diminishes in intensity (see blue tick mark that tracks the VBM based on EDC fits).  These shifts are summarized in the top panel of Fig.  \ref{fig:1}(d).  The LB-VBM peak-to-peak energy difference $\Delta$, which characterizes the hybridization gap,  is calculated as a function of $n$ from the fits and presented in the bottom panel of Fig.  \ref{fig:1}(d), revealing an increase on the order of 30 meV from depleted to occupied flat bands.  This increasing hybridization gap is an important observation for the potential many-body effects in this material,  because the enhanced energy separation reduces interband interactions.

A detailed analysis of the miniband dispersion within a narrow ($E$,$k$)-window,  demarcated by dotted rectangles at 9 V and 12 V in Fig.  \ref{fig:1}(b),  is presented in Fig.  \ref{fig:1}(e) and Fig. 9.  The spectra have been determined by applying the curvature method to the raw ARPES intensity,  which enhances the faint miniband dispersion \cite{CurvM:2011}.  We identify the intense flat band segment as the minimum of the LB at $\gamma$, and the upwards dispersing segments as the dispersion towards the $\kappa$- and $\kappa^{\prime}$-points.  We find that the UB can not be fully occupied before reaching the saturation point for doping in our device.  We thus merely observe intensity from the UB spilling below the Fermi energy,  touching the LB at $\kappa$- and $\kappa^{\prime}$.  The complete LB dispersion is captured by fitting the hopping paramter $t$ and band offset $E_0$ in a two-band tight binding model for a honeycomb lattice to the peak curvature intensity (see Fig.  9 and Methods).  It has been overlaid as orange curves in Fig.  \ref{fig:1}(e).  The curvature intensity continues uninterrupted into the UB states and we find no indication of a gap formation in the observed filling range. 

We estimate the bandwidth of the UB dispersion by requiring the carrier density that can be calculated from the tight binding model to be identical to the experimentally determined values at a given gate voltage (see Methods for further details). The resulting gate-dependent miniband dispersion is plotted relative to the $\kappa$-point energy in Fig.  \ref{fig:1}(f).  The UB bandwidth is generally larger than that for the LB. Furthermore, we observe that the LB widens while the UB flattens as the doping increases. Both observations are consistent with many-body calculations and experiments \cite{Guinea:2018,Rademaker:2019,Cea:2019,Goodwin:2020,Choi:2021}.  Indeed,  similar many-body effects have been shown via ARPES to drive band flattening of the $\pi$-band in single-layer graphene \cite{Ulstrup:2016}. \\

\begin{figure*}[t!] 
\begin{center}
\includegraphics[width=1.0\textwidth]{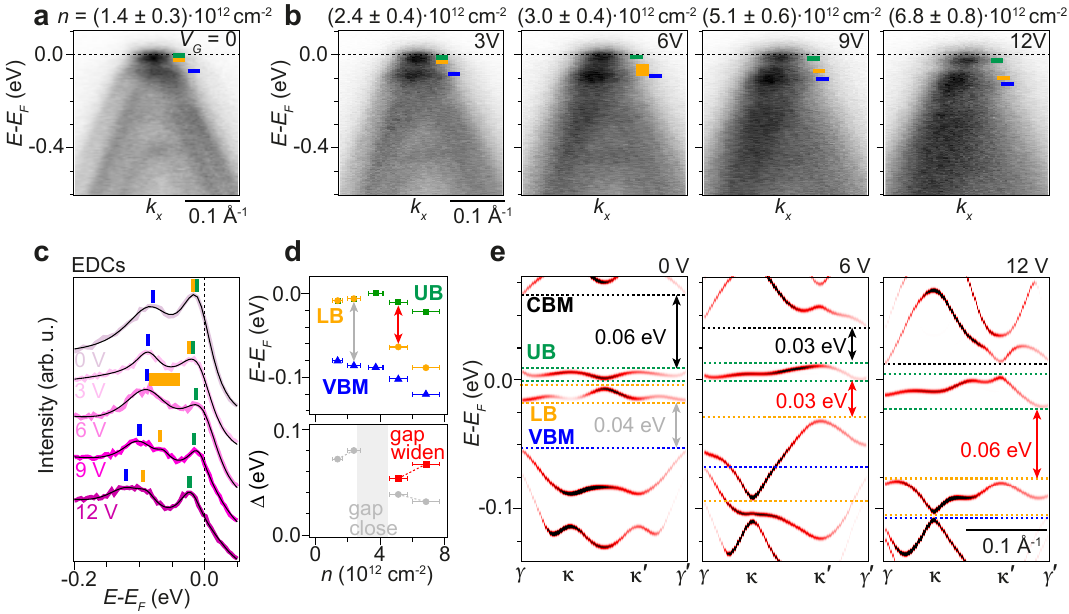}
\vspace{-1cm}
\caption{Gate-tunable electronic structure of TDBG.  \textbf{a-b,} ARPES spectra of TDBG at the stated values of gate voltage and carrier density.  \textbf{c,} EDCs (colored curves) extracted over a $k$-range of 0.04 \AA$^{-1}$ around the flat band region and fits (smooth curves) to Lorentzian peaks multiplied by the Fermi-Dirac function at the stated gate voltages.  Orange,  green and blue ticks demarcate peak positions corresponding to LB, UB and VBM,  respectively.  The peak positions are also indicated in (\textbf{a})-(\textbf{b}).  Note that the LB and UB are indistinguishable at 0 and 3 V.  \textbf{d,} Top panel: Fitted peak positions of UB,  LB and VBM from the EDC analysis.  Bottom panel: Peak-to-peak energy separations between LB and VBM (grey markers),  as well as LB and UB (red markers).  \textbf{e,} Calculated spectral functions at the given $V_G$.  CBM and VBM offsets are marked by horizontal black and blue dashed lines, respectively.  UB and LB are encompassed by horizontal green and orange dashed lines, respectively.  Double-headed black,  grey and red arrows demarcate CBM-UB, LB-VBM and UB-LB band gaps, respectively.  Error bars on the carrier density are described in connection with the doping calibration in Fig.  7.  Error bars derived for EDC peak positions and values of $\Delta$ from the EDC analysis are smaller than the marker sizes. }
\label{fig:2}
\end{center}
\end{figure*}

\textbf{Doping-dependent dispersion of TDBG.}  Figures \ref{fig:2}(a)-(b) present the gate-tunable ARPES dispersion of TDBG.  Similarly as for TBG,  we extract VBM,  LB and UB band positions from EDC analysis in Fig.  \ref{fig:2}(c) and track the fitted positions via blue,  orange and green tick marks,  respectively.  The $n$-dependence of peak energies is presented in the top panel of Fig.  \ref{fig:2}(d).  Peak-to-peak energy separations, shown via grey and red double-headed arrows, represent LB-VBM and UB-LB hybridization gaps,  respectively.  Their values have been plotted in the bottom panel of Fig.  \ref{fig:2}(d).  Initially,  the LB and UB states are not possible to disentangle within the main flat band intensity at the Fermi level.  The fitted EDC peak is therefore understood as being composed of both states at 0 and 3 V.  A hybridization gap on the order of 75 meV is observed towards the VBM in this low-voltage regime.  At 6~V,  we measure intensity that is continuously filling the gapped region towards the VBM.  Upon further doping,  intensity is again depleted from the gapped region and two closely-spaced peaks are observable around -0.1~eV. These are interpreted as the LB and VBM with a strongly suppressed hybridization gap, compared to the initial situation.  A UB-LB hybridization gap on the order of 50~meV is observable at 9 V and increases to 70~meV upon further doping at 12 V.

These observations and in particular the difficulty of observing the LB at a gating voltage of 6~V can be understood by comparison to spectral function calculations within a Coulomb interaction-corrected tight-binding model (see Methods for more details),  as shown in Fig.~\ref{fig:2}(e) and Fig.  10.  The ($E$,$k$)-dependence of the calculated spectral weight and the apparent asymmetry of the bands stem from the sublattice weights in the band unfolding procedure,  as described in Methods.  The calculated spectral function at 0 V reveals two closely spaced LB and UB minibands,  consistent with the ARPES data.  Gaps of 60~meV and 40~meV separate the UB and LB from the CBM and VBM,  respectively.  As the gate voltage is tuned the flat bands separate by up to 60~meV at 12~V and the remaining gaps decrease or vanish entirely.  The LB and UB initially display similar narrow bandwidths.  As the voltage is increased to 6~V,  the LB acquires substantial dispersion and shifts to lower energy while the UB remains pinned around the Fermi energy  The LB then flattens again as it shifts further down upon further increasing the gate voltage.  This behavior matches our experimental observation of intensity filling the gapped region at 6~V, precluding a clear observation of the LB.  We have therefore marked the LB via a widened tick mark at 6 V in Figs.  \ref{fig:2}(b)-(c). The theory also explains the re-emergence of the narrow LB peak at 9~V and 12~V in the ARPES data by the re-flattening of the LB state seen in the calculations.  The substantial band broadening of the LB leads to the LB-VBM gap closing in the intermediate doping range,  as illustrated in the bottom panel of Fig.  \ref{fig:2}(d).  By contrast the UB-LB gap at high doping agrees well with the calculated value because both bands are narrow and the determined peak position thus corresponds well to the band centre.

\section*{Discussion}

\begin{figure*}[t!] 
\begin{center}
\includegraphics[width=0.5\textwidth]{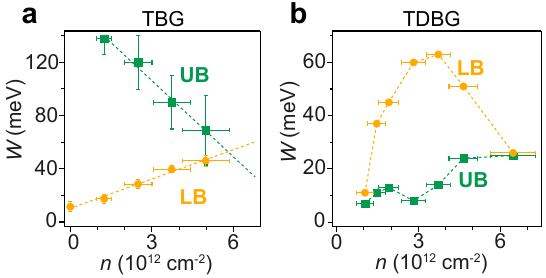}
\vspace{-0.5cm}
\caption{Bandwidth renormalization with doping in TBG and TDBG.  \textbf{a,} LB and UB bandwidths obtained from the fitted TBG dispersions in Fig.  \ref{fig:1}(f).  Dashed lines are linear fits that have been added as a guide to the eye.  \textbf{b,} LB and UB bandwidths of TDBG extracted from the calculated spectral functions in Figs.  \ref{fig:2}(e) and Fig.  10.  The error bars for the LB bandwidth in TBG have been propagated from the dispersion fits in Fig.  9,  while the error bars on the UB bandwidth are propagated from the LB bandwidth and $n$.  For TDBG the error bars on W are associated with how accurate we can determine the band edges,  which leads to error bars smaller than the data points.  The error bars on $n$ are derived from the calibrations shown in Figs.  6-7.}
\label{fig:3}
\end{center}
\end{figure*}

Our findings underline the vast design freedom in moir\'e systems by revealing a striking difference in the doping-dependent band dispersion of TBG and TDBG. The behaviour of TBG is non-trivial in the sense that the minibands experience a non-rigid shift upon increased filling, corresponding to a monotonous bandwidth change of the LB and UB in opposite directions (see  Fig.  \ref{fig:3}(a)). However, this is easily captured by a simple tight-binding model with a variable hopping parameter $t$. Moreover, the results are consistent with the theoretically expected behaviour \cite{Guinea:2018,Rademaker:2019,Goodwin:2020} and previous experiments on single layer graphene \cite{Ulstrup:2016}. By contrast, the filling-dependent band dispersion observed in TDBG is highly complex.  The LB and UB bandwidth changes determined from the calculated spectral function are summarized in Fig.  \ref{fig:3}(b).  Upon increasing the filling, all bands shift to lower energies but they also drastically change their dispersion with a non-monotonous bandwidth change of the LB.  The $n$-dependence of the bandwidth ultimately influences the $U/W$ ratio in the LB and UB that leads to Mott insulating and superconducting phases at low temperatures.  At doping levels where the correlated insulator phase emerges in the LB we find bandwidths in the range of 10-20 meV for both TBG and TDBG.  In the doping regime where correlation effects emerge in the UB,  we estimate somewhat larger bandwidths,  on the order of 60 meV and 20 meV for TBG and TDBG,  respectively.  These values reflect the complex doping-dependent asymmetries between the flat band dispersions that underpin the rich temperature- and doping-dependent electronic phase diagrams in these materials \cite{Polshyn:2019}.

\section*{Methods}

\textbf{Fabrication of devices.} Stacking of 2D flakes was accomplished using the dry-transfer technique applying polymers \cite{Kyounghwan:2016, Purdie2018}. Graphene and hBN layers were exfoliated onto SiO$_{2}$/Si substrates. Their size and thickness were identified by optical microscopy  and AFM.  Bilayer and monolayer graphene flakes were cut in two pieces by using a sharp tip with 2 $\mu$m diameter. One of the pieces was picked up using a polycarbonate (PC)/polydimethylsiloxane (PDMS) stamp mounted on a glass slide at 130 $^{\circ}$C. The remaining piece was rotated by an angle $\theta$ \cite{Kyounghwan:2017}, which was set $0.1^{\circ}$ higher than the target angle and stacked with the mono- or bilayer graphene flake on the stamp in order to form twisted bilayer graphene or twisted double bilayer graphene. Then,  a hBN flake with thickness around $30$ nm was picked up at 120 $^{\circ}$C, which was followed by picking up a graphite flake at  130 $^{\circ}$C to be used as a back gate.  Finally,  the stack was released onto SiO$_{2}$/Si with Au/Cr-patterned electrodes based on $50$ nm gold and $5$ nm chromium films at 180 $^{\circ}$C.  Twisted graphene layers and back gates were contacted by different electrodes.  Heterostructures were then annealed in a H$_{2}$/Ar environment at 250 $^{\circ}$C for several hours prior to measurement.\\

\textbf{C-AFM Measurements.} Conductive tip (C-AFM) measurements were performed in inert N$_2$ atmosphere ($<1$ ppm H$_2$O and O$_2$) using a commercial Asylum Research Cypher S AFM integrated into a Jacomex GP(Concept) glove box. To preserve the twist angles, samples were measured immediately after performing microARPES measurements, without any further annealing or cleaning treatments.  C-AFM scans were performed using doped-diamond-coated conductive tips (NanoSensors$^{\mathrm{TM}}$ CDT-CONTR).  A bias voltage was applied to the sample via a series resistance of 1 G$\Omega$ to limit the current through the sample. The average periodicity of the moir\'e patterns was extracted using Gwyddion software and determined to be $\theta=(1.2 \pm 0.2)^{\circ}$.\\

\textbf{microARPES measurements.} The microARPES measurements were performed at the SGM4 beamline of the ASTRID2 light source at Aarhus University, Denmark \cite{Volckaert:2023}. The devices were wirebonded to CSB00815 chip carriers and annealed in the microARPES end-station at a temperature of 180 $\degree$C for 1~hour before exposure to the beam. Throughout the data collection, the samples were kept at room temperature in a base pressure better than $2\cdot 10^{-10}$~mbar.

The synchrotron beam with photon energy of 47~eV was focused to a spot size of 4~$\mu$m using an elliptical capillary mirror [Sigray Inc.]. The ARPES spectra were obtained using a SPECS Phoibos 150 SAL analyser. The energy- and angular resolution were better than 20~meV and 0.1$\degree$, respectively. The scans of $(E, k_x, k_y)$-dependent photoemission intensity were acquired with the scanning angle lens mode of the analyser while keeping the sample position fixed. The $(E, k, x, y)$-dependent datasets were obtained by rastering the sample position relative to the beam using a piezoelectric stage [SmarAct] and measuring an $(E, k)$-dependent snapshot at each point. The samples were aligned along the $\kappa$-$\kappa^{\prime}$ direction of the mBZ.  Electrostatic doping of the devices in situ was achieved using Keithley 2450 source meters. 

The miniband dispersion of TBG was obtained by fitting the curvature intensity in Fig.  9 to a two-band tight binding model on a honeycomb lattice given by $E_{LB(UB)}(k) = \alpha_{LB(UB)}t_{LB(UB)}\sqrt{3+2\cos(\lambda_s k) + 4\cos(\lambda_s k/2)}-E_0$ where $t_{LB(UB)}$ is the LB (UB) hopping parameter,  $\alpha_{LB}=-1$,  $\alpha_{UB}=1$, $E_0$ is the band offset due to doping and  $\lambda_s = a/2\sin(\theta/2)$.  The value of $\lambda_s$ was fixed according to the twist angle measured by C-AFM.  The bandwidth of the upper flat band is constrained by the requirement that the integral $\int g(E,W)n_{FD}(E)dE$ should return the measured carrier density $n$.  Here, $g(E,W)$ is the density of states of the fitted tight binding bands with bandwidth $W$ and $n_{FD}$ is the Fermi-Dirac function at room temperature.\\

\textbf{Spectral function calculations for TDBG.} Fully relaxed crystal structure of TDBG were obtained on the molecular dynamics simulation using \textsc{LAMMPS}~\cite{thompson_lammps_2022} software with REBO2~\cite{brenner_second-generation_2002} and exact exchange random phase approximation~(EXX-RPA)-fitted DRIP potentials~\cite{leconte_relaxation_2022}. The relaxation was performed for a $1.35^{\circ}$ twisted commensurate cell made with two AB stacked bilayer graphene with a lattice constant of $a=2.46 \AA$. We performed calculations for 14,408 carbon $\pi$-orbitals within the relaxed structure of moiré lattice length of $104.39$ $\AA$.

We employed the following self-consistent interaction corrected tight-binding Hamiltonian to prevent double-counting of the coulomb interactions~\cite{lee_extended_2024}.
\begin{equation}
H = H^{0} +\sum_{i\sigma} (U \Delta \rho_{ii\tilde\sigma} + \sum_j V_{ij} \Delta \rho_{jj}) c^\dagger_{i\sigma} c_{i\sigma}.
\end{equation}
Here, $\Delta \rho$ denotes the disparity between the density matrices of $H$ and the non-corrected Hamiltonian $H^0= \sum_{ij\sigma} t_{ij}c^\dagger_{i\sigma} c_{j\sigma} $. The operator $c_{i\sigma}$($c^\dagger_{i\sigma}$) is electron annihilation(creation) operators, where $\sigma$ denotes the spin index, and $\tilde\sigma$ signifies the spin in the opposite direction. The hopping parameters $t_{ij}$ were determined using the scaled hybrid exponential (SHE) model, which incorporates the effects of structure relaxation~\cite{leconte_relaxation_2022}. We assumed the onsite Hubbard parameter $U = 5$ eV. For intersite interactions $V_{ij}$, values of $V_1 = 2$~eV for first nearest neighbors and $V_2 = 1$~eV for second nearest neighbors were assumed in intralayer relations, while for interlayer relations, the Coulomb equation accounting for bonding length was employed, as $V_{ij} = 1/(\epsilon_r \sqrt{(a/2\sqrt{3})^2 + r_{ij}^2})$ with the relative permittivity $\epsilon_r = 6$~\cite{jung_enhancement_2011}. The exchange interaction was disregarded to simplify the calculations, hence the calculations were carried out at the self-consistent Hartree level. Each gate voltage was considered to be applied to tDBG on 30nm hBN, with saw-like external potential and doping corresponding to the same dielectric constant. The Fermi level of the calculations has been matched to the experiment by comparing the integrated spectral weight with the Lorentzian peaks obtained from the EDC analysis in Fig.~\ref{fig:2}(c),  as shown in Fig.  11.  The self-consistent calculations were performed with $5 \times 5$ grid k-points.

In this calculation, we used the band unfolding approach of Ref.~\cite{nishi_band-unfolding_2017} to calculate the spectral function of moiré supercell systems. The spectral function at a point $k$ of graphene Brillouin zone unfolded from a $K$ point of the moiré Brillouin zone can be represented through
\begin{equation}
A(k,E) = -\frac{1}{\pi}\frac{S_g}{S_m}\sum_{K,G,N}\delta_{k-G,K}\sum_s \omega^s_{KN}(k) \operatorname{Im}[{\frac{1}{E-\epsilon_{KN}+i\eta}}]
\end{equation}
where $G$ is the reciprocal lattice vector of the moiré structure, $N$ is the band index of the moiré band, $s$ denotes the sublattices and the parameter $\eta$ is a broadening term that accounts for energy uncertainty. The $S_g$ and $S_m$ represent the area of graphene unit cell and the unit cell of the moiré pattern, respectively. 
The weight at each sublattice $\omega^s_{KN}(k,E)$ is
\begin{equation}
\omega^s_{KN}(k) = \sum_{I\in s, J \in s} \delta_{i(I),i'(J)} (C^{KN}_I)^*C^{KN}_J e^{ik(T_s(I)-T_s(J))}
\end{equation}
where $I,J$ are the basis indices in the moiré structure, the $i(I)$ labels the basis in the primitive cell of the sublattice redefined from $I$ basis, and the $C^{KN}_I$ is the eigenvector component. The vector $T_s(I)$ is the displacement of the $I$-basis from the primitive cell by a certain number of lattice vectors.

\section*{acknowledgement}
The authors acknowledge funding from the Danish Council for Independent Research, Natural Sciences under the Sapere Aude program (Grant Nos.  DFF-9064-00057B and DFF-6108-00409),  the Aarhus University Research Foundation, the Novo Nordisk Foundation (Project Grant NNF22OC0079960) and from VILLUM FONDEN under the Villum Investigator Program (Grant. No. 25931).  D. L. acknowledges support by the Korean Ministry of Land, Infrastructure and Transport (MOLIT) from the Innovative Talent Education Program for Smart Cities. J. J. acknowledges support from the Korean NRF grant No. NRF2020R1A5A1016518. 
C.S.  acknowledges Marie Sklodowska-Curie Postdoctoral Fellowship (proposal number 101059528).  K.W. and T.T. acknowledge support from the JSPS KAKENHI (Grant Numbers 21H05233 and 23H02052) and World Premier International Research Center Initiative (WPI), MEXT, Japan.  We acknowledge the Urban Big Data and AI Institute of the University of Seoul supercomputing resources (http://ubai.uos.ac.kr) and KISTI Grant No. KSC-2022-CRE-0514 made available for conducting the research reported in this paper.

\pagebreak

\section{Supporting Information}

\begin{figure*} [h!]
	\begin{center}
		\includegraphics[width=1\textwidth]{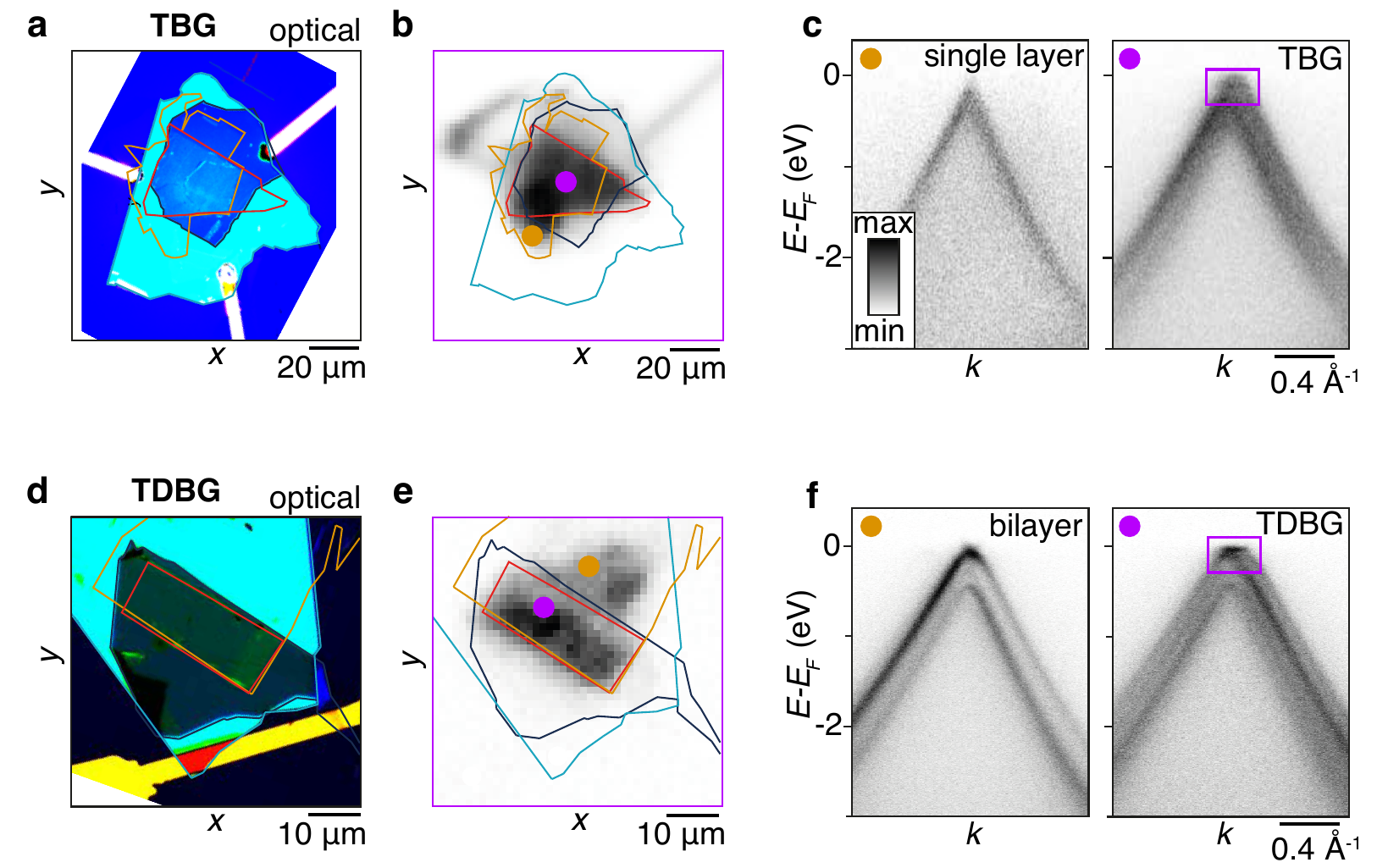}
		\caption{Correlating photoemission intensity with optical images of (\textbf{a})-(\textbf{c}) TBG and (\textbf{d})-(\textbf{f}) TDBG devices.  \textbf{a,} (\textbf{d,})  Optical image of TBG (TDBG) device.  \textbf{b,} (\textbf{e,}) Photoemission intensity map of device composed from the ($E$,$k$)-integrated intensity from flat dispersion segments in TBG (TDBG).  Top and bottom single layer graphene (bilayer graphene) are shown via red and orange outlines, respectively,  while graphite and hBN flakes are demarcated by dark and light blue outlines.  \textbf{c,} (\textbf{f,}) ($E$,$k$)-dependent photoemission intensity for the positions labelled by correspondingly colored circles in the map.  The purple box outlines the ($E$,$k$)-integration region that the map has been composed from.}
		\label{fig:ex1}
	\end{center}
\end{figure*}

\begin{figure*} [t!]
	\begin{center}
		\includegraphics[width=0.9\textwidth]{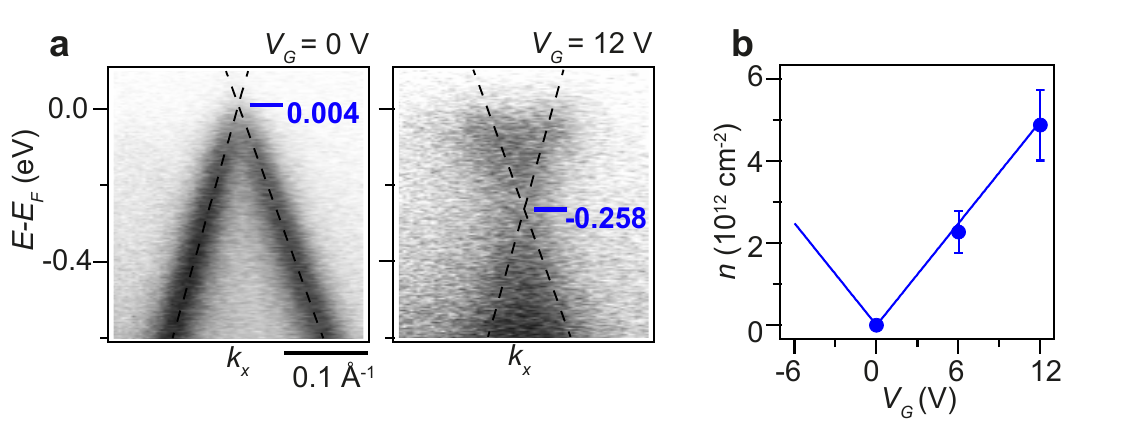}
	 	\vspace{-0.4cm}
		\caption{Calibration of carrier density with gate voltage determined on bottom single-layer graphene area in TBG device.  \textbf{a,} Photoemission intensity around the graphene $\mathrm{K}$-point at 0 V and 12 V,  showing the effect of maximally gating single-layer graphene in our device.  The overlaid dashed linear bands are based on fits to the dispersion that have been extrapolated to estimate the Dirac point energy.  The slope determined at 0 V is kept fixed for the fit at 12 V.  The estimated energy of the Dirac point is demarcated by blue ticks and stated in units of eV. The standard deviation error bars amount to $\pm$0.012 eV and $\pm$0.026 eV at 0 V and 12 V, respectively.  \textbf{b,} Carrier concentration $n$ calculated from the Dirac point energies.  Error bars have been propagated from the standard deviation associated with the Dirac point energy fits. The v-shaped curve is a fit to the expression $a_G|V_G - V_0|$,  where $a_G$ is the slope and $V_0$ is the gate voltage that corresponds to charge neutrality. }
		\label{fig:ex3}
	\end{center}
\end{figure*}

\begin{figure*} [t!]
	\begin{center}
		\includegraphics[width=0.9\textwidth]{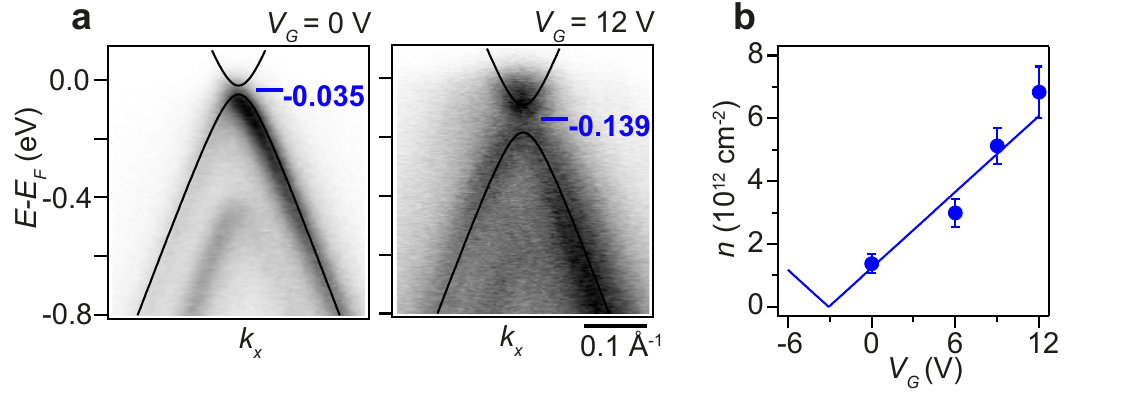}
		\vspace{-0.4cm}
		\caption{Calibration of carrier density with gate voltage determined on bottom bilayer graphene area in TDBG device.  \textbf{a,} Photoemission intensity around the bilayer graphene $\mathrm{K}$-point at 0 V and 12 V,  presenting the effect of maximal gating in this device.  The overlaid hyperbolic bands are based on fits of the dispersion to a two-band massive Dirac fermion dispersion.  The energy of the charge neutrality point (i.e mid-gap energy) resulting from the fits are demarcated by blue ticks and stated in units of eV. The error bars are $\pm$0.007~eV and $\pm$0.013~eV at 0 V and 12 V, respectively.  \textbf{b,} Carrier concentration $n$ with error bars determined from the fits of the charge neutrality point energy.  The v-shaped curve is a fit to the function $a_G|V_G - V_0|$,  where $a_G$ is the slope and $V_0$ is the gate voltage that corresponds to charge neutrality. }
		\label{fig:ex4}
	\end{center}
\end{figure*}

\begin{figure*} [t!]
	\begin{center}
		\includegraphics[width=1\textwidth]{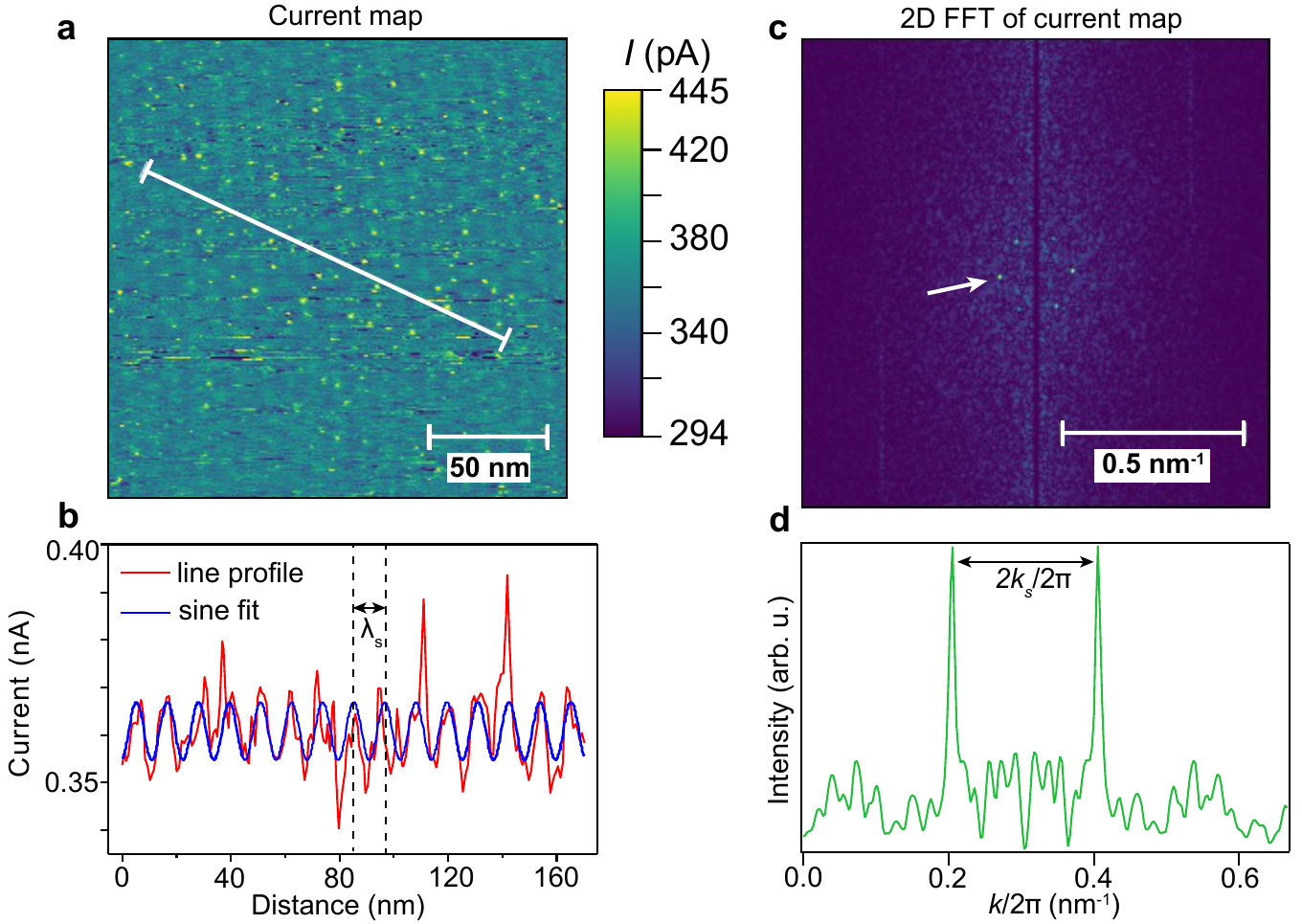}
		\caption{Conductive tip atomic force microscopy (C-AFM) experiment to determine twist angle. \textbf{a,} C-AFM current map of near magic angle TDBG.  The twist angle in TBG is checked using the same method (not shown here). \textbf{b,} Line profile and sine fit extracted along the white line in \textbf{a},  providing an estimate of the moir\'e unit cell periodicity $\lambda_s = a/2\sin(\theta/2)$,  where $a$ is the graphene lattice parameter. \textbf{c,} Reciprocal lattice obtained by fast Fourier transform (FFT) of the image in (\textbf{a}). The arrow indicates a moir\'e reciprocal lattice point. \textbf{d,} Intensity extracted across the hexagonal reciprocal lattice in  (\textbf{c}) with the indicated spacing between the points.  From $k_s$ we obtain $\lambda_s$.  The two methods yield an average twist angle of 1.2$^{\circ}$ with an uncertainty of $\pm 0.2^{\circ}$.}
		\label{fig:ex5}
	\end{center}
\end{figure*}

\begin{figure*} [t!]
	\begin{center}
		\includegraphics[width=1\textwidth]{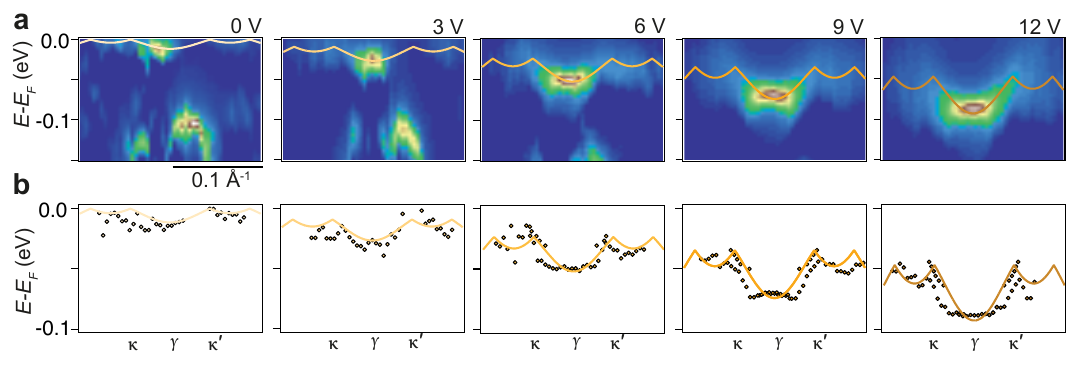}
			\vspace{-1cm}
		\caption{Extraction of miniband dispersion in TBG.  \textbf{a,} Curvature plots of the ARPES intensity at the stated gate voltages.  The fitted dispersion of the LB has been overlaid as light-to-dark orange curves from 0 V to 12 V. \textbf{b,} Fitted peak positions (circles) based on the curvature plots in (\textbf{a}) obtained from combined energy and momentum distribution curve fits.  The miniband dispersion is obtained by fitting the hopping parameter $t_{LB}$ and band offset $E_0$ in the honeycomb tight binding dispersion for the LB given by $E(k) = -t_{LB}\sqrt{3+2\cos(\lambda_s k) + 4\cos(\lambda_s k/2)}-E_0$ in each case where $\lambda_s$ is fixed to the experimental value from C-AFM.}
		\label{fig:ex6}
	\end{center}
\end{figure*}

\begin{figure*} [t!]
	\begin{center}
		\includegraphics[width=1\textwidth]{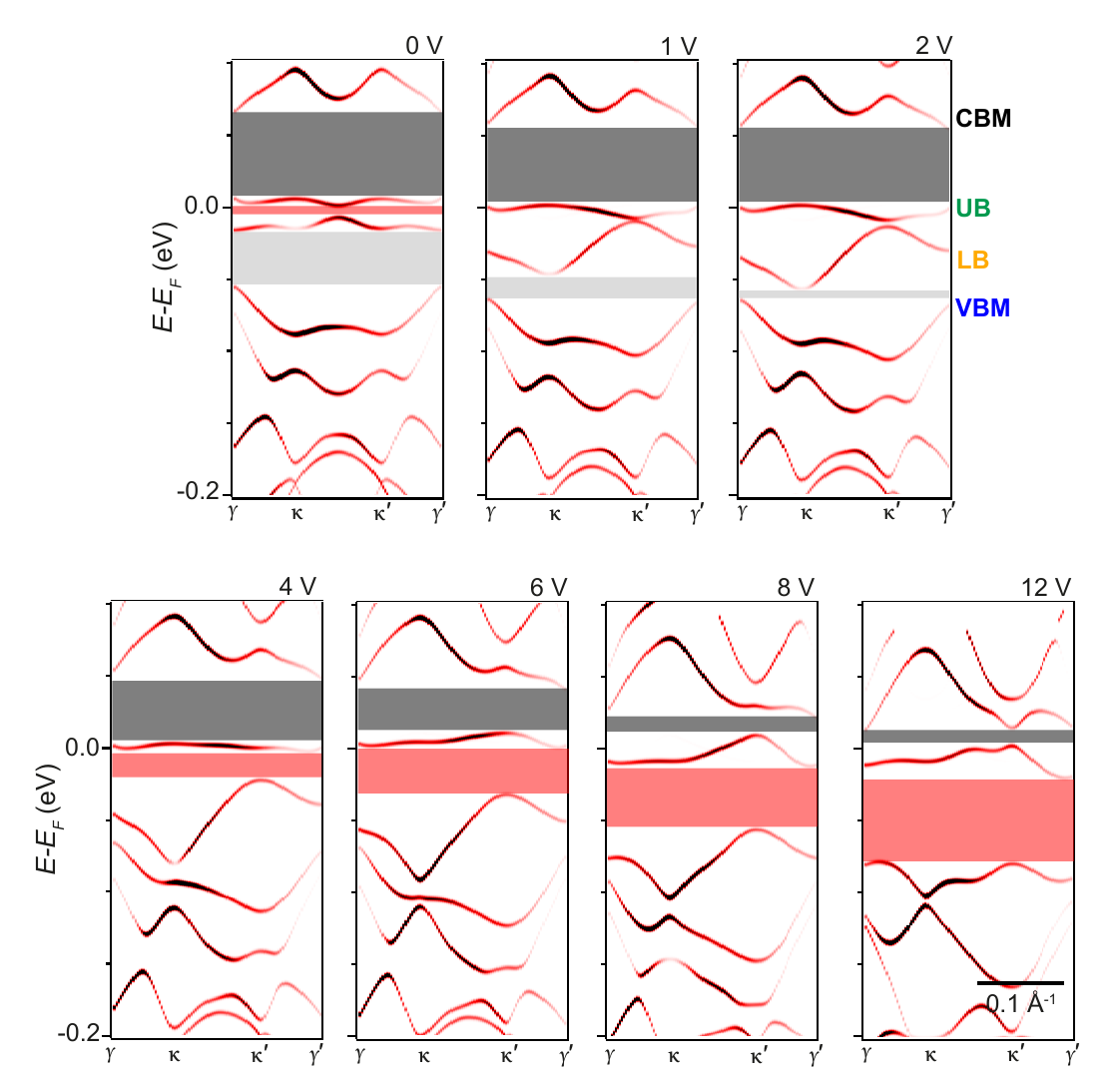}
		\caption{Calculated TDBG spectral functions from 0 V to 12 V.  CBM-UB,  UB-LB and LB-VBM gaps are demarcated by dark grey, red and light grey boxes.}
		\label{fig:ex7}
	\end{center}
\end{figure*}

\begin{figure*} [t!]
	\begin{center}
		\includegraphics[width=1\textwidth]{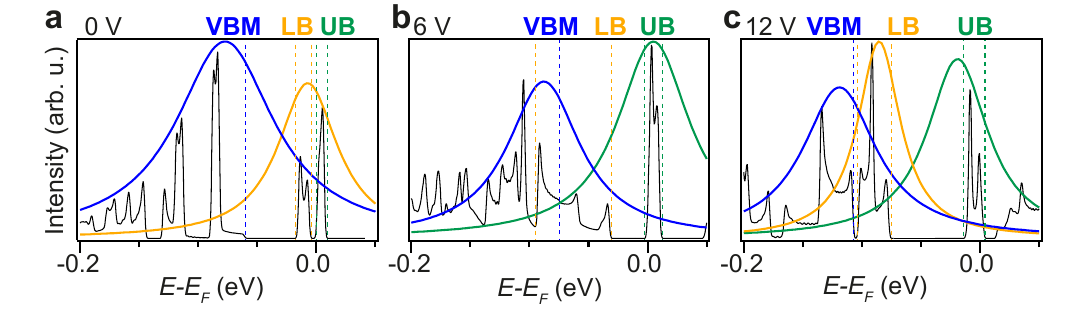}
		\caption{Comparison of measured and calculated spectral weights of TDBG dispersion.  \textbf{a-c,} Lorentzian peaks corresponding to VBM (blue curve),  LB (orange curve) and UB (green curve) states determined from the EDC analysis in Fig.  3(c) of the main paper compared with the integrated spectral weight from the calculations at the given gate voltages. The Fermi energy of the calculated spectral function has been adjusted in each case such that we obtain the best fit between the experimental and calculated peak positions.  This procedure shows that the experimental and calculated band alignments are in agreement,  as peaks in the calculated spectral weight align closely with the Lorentzians obtained from the experiment.  At 6 V,  the LB peak vanishes in both experiment and theory because the LB acquires substantial dispersion and distributes its spectral weight across a large segment of the UB-VBM gap.  At 12 V,  the LB flattens again,  giving rise to a sharp peak in both theory and experiment.  The vertical orange and green dashed lines encompass the LB and UB states, respectively,  while the vertical blue lines demarcate the VBM,  as determined from the dispersion plots in Fig.  10. }
		\label{fig:ex8}
	\end{center}
\end{figure*}

\end{document}